\documentclass[preprint,aps,floats,showpacs,amssymb,tightenlines]{revtex4}
\usepackage{epsfig}

\newcommand{\bml}{\begin{mathletters}}
\newcommand{\eml}{\end{mathletters}}

\newcommand{\bea}{\begin{eqnarray}}
\newcommand{\eea}{\end{eqnarray}}
\newcommand{\be}{\begin{equation}}
\newcommand{\ee}{\end{equation}}
\newcommand{\beast}{\begin{eqnarray*}}
\newcommand{\eeast}{\end{eqnarray*}}
\newcommand{\pkt}{\; .}
\newcommand{\kma}{\; ,}

\def\e{{\rm e}}

\begin{document}




\title{Euclidean solutions of Yang-Mills-dilaton theory}
\renewcommand{\thefootnote}{\fnsymbol{footnote}}

\author{Yves Brihaye \footnote{Brihaye@umh.ac.be}}
\affiliation{Dep de Math\'ematiques et Physique Th\'eorique,
Universit\'e de Mons, Place du Parc, 7900 Mons, Belgique}
\author{George Lavrelashvili \footnote{lavrela@itp.unibe.ch }}
\affiliation{
Department of Theoretical Physics,\\
A.Razmadze Mathematical Institute,\\
GE-0193 Tbilisi, Georgia}
\date{\today}
\setlength{\footnotesep}{0.5\footnotesep}


\begin{abstract}
Classical solutions of the Yang-Mills-dilaton theory in Euclidean space-time are investigated.
Our analytical and numerical results imply existence of infinite number of
branches of dyonic type solutions labelled by the number of nodes of gauge field amplitude $W$.
We find that the branches of solutions exist in finite region of parameter space
and discuss this issue in detail in different dilaton field normalization.
\end{abstract}

\pacs{11.27.+d, 11.15Kc, 04.20.Jb}
\maketitle
\section{Introduction} \label{intro}

Bartnik and McKinnon (BK) discovery of static, spherically symmetric,
globally regular solutions of the Einstein-Yang-Mills (EYM) theory \cite{bm88}
has opened new avenue of research in General Relativity.
Soon  after BK discovery related (colored) black holes were found and different generalisations
were investigated (see e.g. \cite{vg} for review  and references).
Interesting modification of pure EYM theory is inclusion of a dilaton field,
which is predicted from many modern theoretical considerations like
string theory and extra dimensions.
So, EYM-dilaton theory was investigated \cite{lm93,biz93b,dg93}
and solitons and black holes similar to BK case were found.
It came as a surprise that even eventually simpler YM-dilaton system
in flat space-time has infinite tower of globally regular
solutions \cite{lm92,biz93a,maison05} similar to the BK case. These solutions
can be labelled by an integer $n$ - number of nodes of gauge field amplitude.
The lowest solutions in EYM and (E)YM-dilaton theory have one node and
the whole tower of solutions turns out to be unstable
with ${\rm n}^{\rm th}$ solution having $2n$ unstable modes:
$n$ in `gravitational' sector and $n$ in `sphaleron' sector \cite{sz90,lm95,vbls95,asch06}.

Note that most of these investigations were performed in four dimensional space-time with
Lorentzian signature. In such case it was found that electric part of non-abelian $SU(2)$ gauge field
should necessarily vanish for asymptotically flat solutions \cite{eg90,bp92}.
Situation is changed in Euclidean sector: electric field here
plays role similar to the Higgs field in Lorentzian sector.
Euclidean version of the EYMD theory was investigated \cite{br06} and solutions
with non-vanishing electric field and non-vanishing non-Abelian charges were found recently.

The aim of present paper is to study Euclidean solutions in the YM-dilaton theory.
The rest of the paper is organized as follows: in next Section we formulate our model,
derive equations of motion and asymptotic behaviour of finite action solutions.
In Sect. III we discuss some special solutions of our system. In Sect. IV we present
our numerical results. Sect. V contains concluding remarks.

\section{The model} \label{ansatz}%
Starting point of our investigation is the Euclidean version of  Yang-Mills theory
coupled to a dilaton field $\phi$:
\be
S_E=\int \left( \frac{1}{2} \partial_\mu \phi \partial^\mu \phi
+ \frac{\e^{2\gamma\phi}}{4 g^2} F_{\mu\nu} F^{\mu\nu}\right)d^4 x \kma
\ee
where $F_{\mu\nu}$ is the non-Abelian gauge field strength and
$\gamma$ and $g$ denote the dilatonic and gauge coupling constants, respectively.
\subsection{The Ansatz and equations}
In this study we will restrict ourselves with the $SU(2)$ gauge group and will be interested in
spherically symmetric solutions.
Following Witten \cite{wit77,fm80} the general spherically symmetric $SU(2)$ Yang-Mills field
can be parameterized as follows:
\be
A_0^a= \frac{x^a}{r} u(r) \kma~\qquad
A_j^a=\epsilon_{jak} x_k \frac{1-W(r)}{r^2}
+\left[\delta_{ja}-\frac{x_j x_a}{r^2}\right]\frac{A_1}{r}
+ \frac{x_j x_a}{r^2} A_2 \kma
\ee
with  $r=\sqrt{x_i^2}$ being the radial coordinate,
$i,j=1,2,3$ are spatial indices and $a,b=1,2,3$ group
indices.

The reduced action $S_{red}$ is defined as follows:
\be
S_E=\int_0^{\tau_{\rm max}} d\tau S_{red} \kma
\ee
with $\tau=x_0$ being Euclidean time. Without loss
of generality we can put $A_1=A_2=0$.
The reduced action then reads:
\be \label{red_act}
S_{red}= \int dr \Bigl( \frac{1}{2} r^2 \phi'^2 +\frac{\e^{2\gamma\phi}}{g^2}
(W'^2+\frac{(1-W^2)^2}{2r^2}+\frac{1}{2} r^2 u'^2 + W^2 u^2) \Bigr) \kma
\ee
where the prime denotes the derivative with respect to $r$.

The equations of motion which follow from the reduced action (\ref{red_act})
read:
\bea \label{eqmW}
W''&=&-2 \gamma \phi' W' -\frac{(1-W^2) W}{r^2} + W u^2 \kma \\ \label{eqmu}
u''&=&-2\frac{u'}{r}-2\gamma \phi' u' +\frac{2}{r^2} W^2 u \kma \\\label{eqmf}
\phi''&=& -\frac{2}{r}\phi'+\frac{2\gamma \e^{2\gamma \phi}}{g^2 r^2}
(W'^2+\frac{(1-W^2)^2}{2 r^2}+\frac{r^2 u'^2}{2}+W^2 u^2 ) \pkt
\eea

The rescaling
\be
\phi \to \frac{\phi}{\gamma}\kma~\qquad r\to \frac{\gamma}{g} r
\kma~\qquad u \to \frac{g}{\gamma} u \kma
\ee
removes the gauge and dilatonic coupling constants $g$ and $\gamma$, respectively,
from the equations of motion. We thus set without loosing generality
$\gamma=g=1$ in the following.

\subsection{Asymptotic behaviour}

Close to $r=0$ we find a 3-parameter family of solutions regular at the origin,
which can be parameterized in terms of $b$, $u_1$ and $\phi_0$:
\bea
\label{zero1}
W(r)&=&1-b r^2 + O(r^4)\kma \\
\label{zero2}
u(r)&=&u_1 r -\frac{u_1}{2} r^2 + O(r^3) \kma  \\
\label{zero3}
\phi(r)&=& \phi_0 + 2 \e^{2\phi_0} \left(b^2+\frac{u_1}{4}\right) r^2+ O(r^3) \pkt
\eea

The behaviour of the solutions at infinity is more involved.
Assuming a power law behaviour for $u$ and $\phi$,
for leading behaviour of $W$ we obtain expressions containing the
Whittaker functions, which in leading order decay exponentially. We thus obtain:
\bea
\label{inf1}
W(r)&=& C \e^{-u_{\infty} r} r^{Q_e}(1+ O(\frac{1}{r})) \kma \\
\label{inf2}
u(r)&=&u_{\infty}-\frac{Q_e}{r} + O(\frac{1}{r^2}) \kma \\
\label{inf3}
\phi(r)&=& \phi_{\infty} -\frac{Q_d}{r}+O(\frac{1}{r^2}) \pkt
\eea
So, at infinity there are five free parameters
$\phi_{\infty}, u_{\infty}, Q_e$, $Q_d$ and $C$.

Note that the shift of dilaton field $\phi \to \phi
+ \tilde{\phi}$ for any {\it finite value}
$\tilde{\phi}$ can be compensated by a rescaling of $r$ and $u$ as follows:
\be
r \to r \e ^{\tilde{\phi} }\kma~\qquad u\to u \e^{-\tilde{\phi}} \pkt
\ee
This symmetry can be used to normalize the solutions
e.g. as $\phi (r=0)=0$ or $\phi (r=\infty)=0$,
and reduces the number of free parameters in the system by one.

\section{Special solutions} \label{analytic}
\subsection{Analytic solutions}
The equations of motion (\ref{eqmW}), (\ref{eqmu}), (\ref{eqmf})
have three important analytic solutions.
The first one is the vacuum solution with
\be \label{vacuum}
W(r)=\pm 1 \ \ , \ \  u(r)=0 \ \ , \ \  \phi(r)=0 \ \ .
\ee

The second one is a simple, dyonic-type generalization of
dilatonic monopole solutions which have been found previously
in the case for $u(r)=0$ \cite{biz93a}:
\be \label{monopole}
W(r)=0 \ \ , \ \  u(r)=\bar{u} \ \ , \ \  \phi(r)={\rm ln} (\frac{r}{1+r}) \ \ ,
\ee
where $\bar{u}$ is an arbitrary constant. As in the pure magnetic case
this solution can be obtained by integrating the first order
Bogomol'ny-type equation
as follows. The reduced action (\ref{red_act}) for $W(r)=0$ and $u(r)=const$
can be written as
\be \label{act}
S_{red}= \frac{1}{2}  \int dr \left( r \phi' - \frac{\e^{\phi}}{r}\right) ^2
+\e^{\phi}|^\infty_0 \pkt
\ee
Thus for configurations with values
$\phi(0)=-\infty$ and $\phi(\infty)=0$ the
action (\ref{act}) has a minimum $S_{red}=1$ and
attains this minimum for solutions
of the first order Bogomol'ny-type equation
\be \label{bog}
r^2 \phi'- \e^{\phi}=0 \pkt
\ee
Solutions of this equation automatically satisfy
the equations of motion (\ref{eqmW})-(\ref{eqmf})
with $W=0$ and $u=const$.
Integrating  (\ref{bog}) we obtain the
logarithmic behavior (\ref{monopole}) for the dilaton field.

The third solution is
well known
Bogomol'ny-Prasad-Sommerfield (BPS)
monopole solution, which in our notations reads:
\be \label{bps}
W=\frac{r}{{\rm sinh}(r)} \kma~\quad u={\rm coth}(r)-\frac{1}{r} \kma~\quad \phi=0 \pkt
\ee

\subsection{Solutions for $u=0$}
The system of Eqs.~(\ref{eqmW}-\ref{eqmf}) for $u\equiv 0$ was first analyzed in \cite{lm92}
and it was shown that a series of solutions exist
indexed by the number of nodes $n$ (with $n \geq 1$) of the function $W$.
In the normalisation $\phi(0)=0$ each of these solutions can be uniquely characterized by a value
of the `shooting' parameter $b$ entering in (\ref{zero1}); for the first few  $n$ we have
\be
     b_1 \approx 0.26083  \ \ , \ \ b_2 \approx 0.35352 \ \ , \ \ b_3 \approx 0.37500 \pkt
\ee
The precise numerical values of parameters $b_n$ for
$n=1,..,8$  can be found  in \cite{maison05}.
The solutions with a higher number of nodes
are very well approximated on an increasing domain of $r$ by
\begin{equation} \label{lim_stat}
W(r)= \frac{\tilde{W_1}}{\sqrt{r}} {\rm sin}\left(\frac{\sqrt{3}}{2}
\ln r +\delta\right) \ \ , \ \ u(r)=0  \ \ , \ \ \phi(r)=\ln r \kma
\end{equation}
which is solution of the linearized equation of $W$
\be
    W'' + \frac{2}{r} W' + \frac{1}{r^2}W = 0 \kma
\ee
with $\tilde{W_1}$ and $\delta$ arbitrary constants.
The limiting solution with infinitely many zeros of $W$ described by the
asymptotic behavior (\ref{lim_stat})
has $b\to b_{\infty}\approx 0.37949$.

The values of masses (reduced action) of static solutions in the normalisation $\phi(\infty)=0$
lie between $M_1\approx 0.80381$
for the solution with one node and $M_{\infty} =1$ for the limiting solution \cite{lm92,maison05}.

\subsection{Related Bessel function}
In the case of a non-zero electric field, the nature of limiting solution is changed.
For
\be
u = q \ \  {\rm and}   \ \ \phi = \ln r \kma
\ee
for the gauge amplitude $W$ in the linear regime, $W<<1$, we get equation
\be \label{bessel}
    W'' + \frac{2}{r}W'+ ( \frac{1}{r^2} - q^2)W = 0 \ \ ,
\ee
whose  solution  is a spherical Bessel function
\be\label{lim_sol}
W(r)= W_{\rm bessel} \equiv \frac{\tilde{W}_2}{\sqrt{r}} {\rm K}_{\frac{i\sqrt{3}}{2}}(q r) \ \ , \ \
\ee
where   $\tilde W_2$ is an arbitrary constant.

\begin{figure}[!htb]
\centering
\leavevmode\epsfxsize=10.0cm
\epsfbox{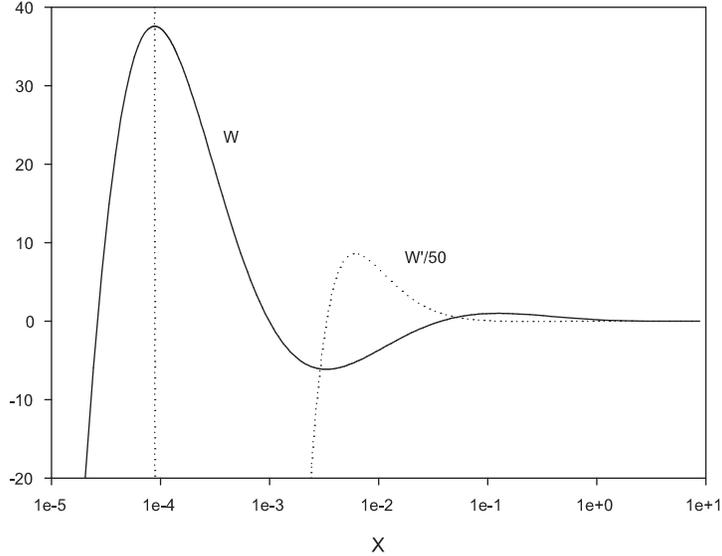}\\
\caption{\label{bessel_fig}
The solution of the Bessel equation (\ref{bessel}).
$W$ and $W'$ are plotted as functions of $x$.}
\end{figure}

The  function  $W_{\rm bessel}$  has the following properties, where
for convenience we use $x = qr$.
For $x \gg  1$ the solution decays according to $W = \exp(-x)(1/x + 0(1/x^2))$. Close to the
origin the solution is not analytic. Solving the equation numerically, we see that
the for $ x < 1$ the solution develops oscillations whose amplitudes become larger while the limit $x \to 0$
is approached. This is illustrated on Fig. \ref{bessel_fig}.
The first nodes appear for
\be
x_1 \approx 0.0367 \ \  , \ \ x_2 \approx 0.001 \ \ , \ \   x_3 \approx 0.000026.
\ee
These zeros are separated by extrema, alternatively maxima and minima. The first of these points  appear at
\be
\tilde x_1 \approx 0.1265 \ \ , \ \ \tilde x_2 \approx 0.0032 \ \ , \ \ \tilde x_3 \approx 0.0008.
\ee
where, up to an arbitrary normalisation, $W$ takes values
\be \label{bessel_extrema}
   W(\tilde x_1) = 1 \ \ , \ \ W(\tilde x_2) \approx -6.12 \ \ , \ \ W(\tilde x_2) \approx 35.56
\ee

\section{Numerical results} \label{num_res}
\subsection{Solutions in the gauge $\phi(\infty)=0$ }
We have solved the above equations numerically using the
differential equations solver COLSYS \cite{colsys}.
To construct these solutions, we use the following boundary
conditions:
\begin{equation} \label{bczero}
W(0)=1 \ , \ u(0)=0 \ , \  \phi(r)'\vert_{r=0}=0
\end{equation}
at the origin and
\begin{equation} \label{bcinfty}
W(\infty)=0 \ , \ u(\infty) =
u_{\infty} \ , \ \phi(\infty)=0
\end{equation}
at infinity.
The parameters $b^{res}$, $u_1^{res}$, $u_{\infty}^{res}$ and mass $M^{res}$
for the solutions in the normalization $\phi_{\infty}=0$ are related to the parameters
$b$, $u_1$, $u_{\infty}$ and mass $M$ in the normalisation $\phi_0=0$
with a simple shift and rescaling:
\be
\label{res}
\phi_0^{res}=-\phi_{\infty} \kma
b^{res}=\e^{2\phi_{\infty}} b \kma
u_1^{res}=\e^{2\phi_{\infty}} u_1 \kma
u_{\infty}^{res}= \e^{\phi_{\infty}} u_{\infty} \kma
M^{res}=\e^{-\phi_{\infty}} M \pkt
\ee

Practically, $u(\infty)$ is the parameter that is fixed by hand in the
boundary conditions. From a numerical solution with fixed $n$ and $u(\infty)$
one can extract the values of $u_1$ and $b$, and other parameters which enter in asymptotic
behaviour Eqs.~(\ref{zero1}-\ref{zero3}) and Eqs.~(\ref{inf1}-\ref{inf3}).

\begin{figure}[!htb]
\centering
\leavevmode\epsfxsize=9.0cm
\epsfbox{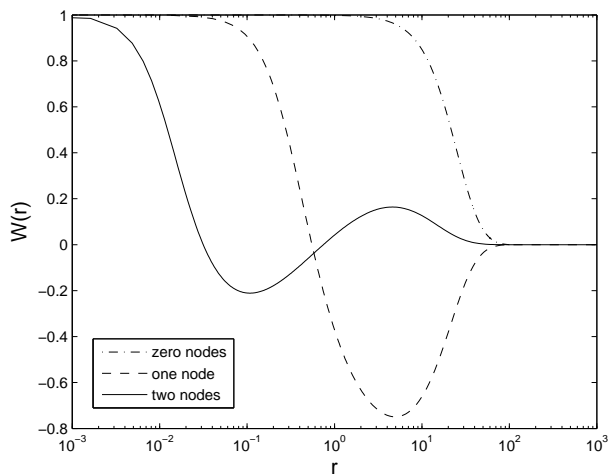}\\
\caption{\label{fig1}
The gauge field function $W$ is shown for zero,
one and two node-solutions, respectively, for $u^{res}(\infty)=0.1$.}
\end{figure}

\begin{figure}[!htb]
\centering
\leavevmode\epsfxsize=9.0cm
\epsfbox{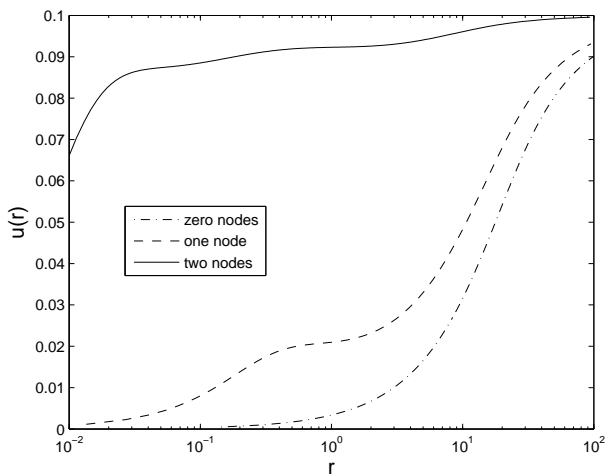}\\
\caption{\label{fig1b}
Same as Fig.\ref{fig1} for the gauge field function $u^{res}$.}
\end{figure}

\begin{figure}[!htb]
\centering
\leavevmode\epsfxsize=9.0cm
\epsfbox{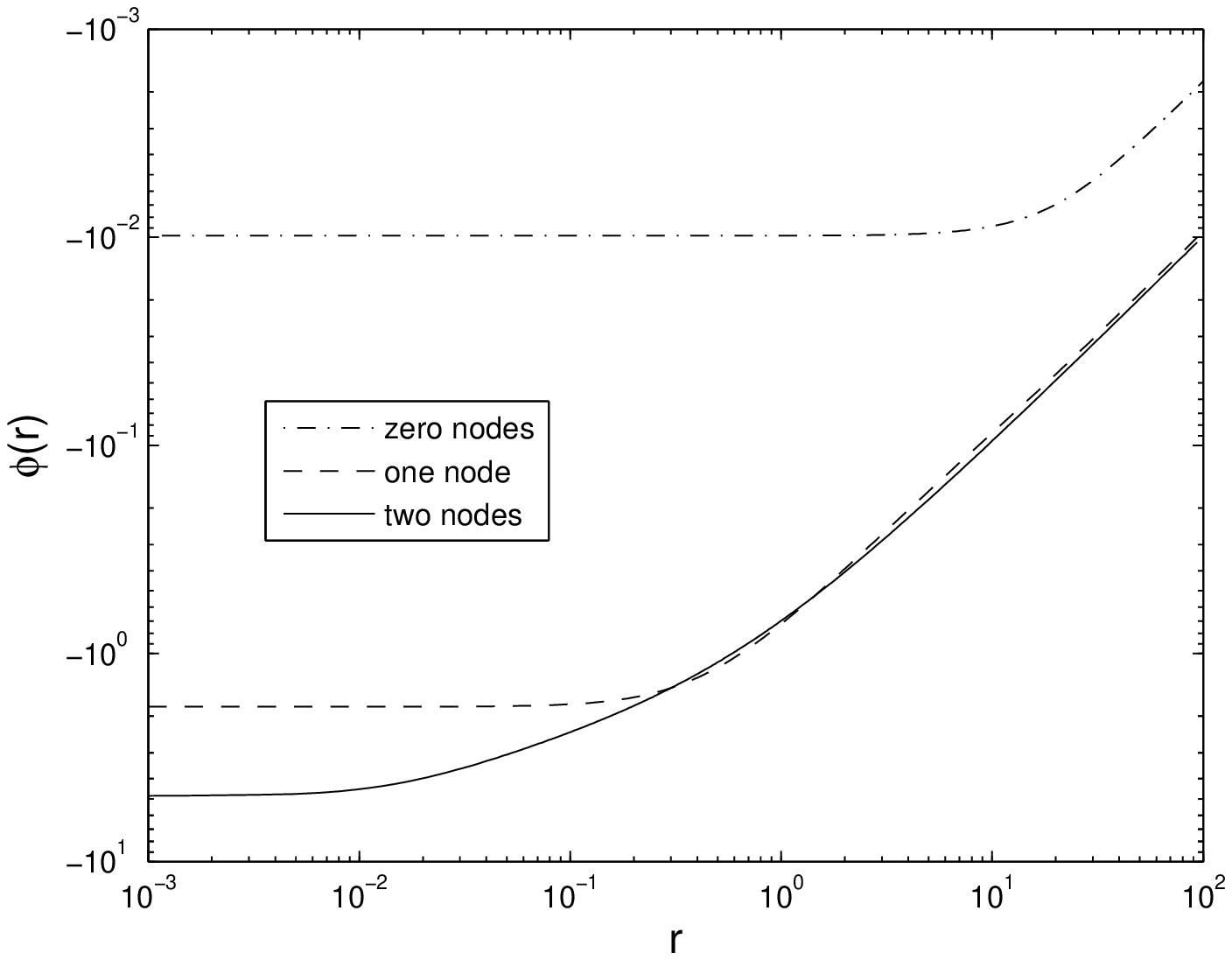}\\
\caption{\label{fig1c}
Same as Fig.\ref{fig1} for the dilaton field function $\phi^{res}$.}
\end{figure}

We studied in details the solutions of the above equations with zero,
one and two nodes of the gauge field function $W(r)$.
Examples of such solutions are given in Fig. \ref{fig1}-\ref{fig1c}.
for $u^{res}(\infty)=0.1$ respectively for the gauge field function $W(r)$
for zero, one and two nodes, and the corresponding profiles for the functions $u$ and $\phi$.

\subsection{Pattern of the solutions}
In order to relate the pattern of solutions available for $u \neq 0$ to the solutions of
\cite{lm92},\cite{maison05} we convert our
solutions to the normalization $\phi_0=0$ by (\ref{res}).
The values of the two ``shooting parameters''
$b$ and $u_1$ then
fully characterize the solutions and describe a point $P(b,u_1)$ on a plot
of initial data. On the $u_1=0$-axis, the numerical values found for $b$
agree with the solutions of \cite{lm92} and leads to the points  $(b_n,0)$ where
$b_n$ are given with a high accuracy in \cite{lm92},\cite{maison05}.

\begin{figure}[!htb]
\centering
\leavevmode\epsfxsize=9.0cm
\epsfbox{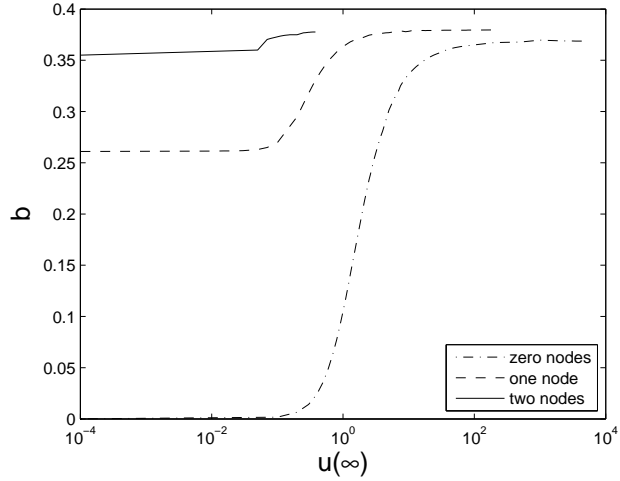}\\
\caption{\label{fig2}
The dependence of the parameter $b$ on $u(\infty)$ is shown
for zero, one and two node solutions.}
\end{figure}

\begin{figure}[!htb]
\centering
\leavevmode\epsfxsize=9.0cm
\epsfbox{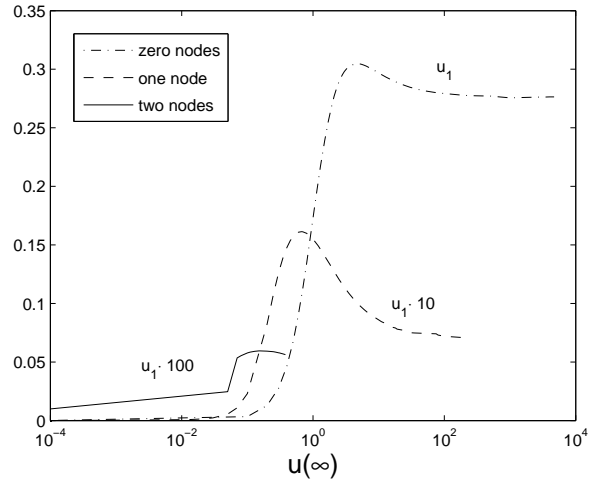}\\
\caption{\label{fig3}
Same as Fig.\ref{fig2} for the parameter $u_1$.}
\end{figure}

\begin{figure}[!htb]
\centering
\leavevmode\epsfxsize=9.0cm
\epsfbox{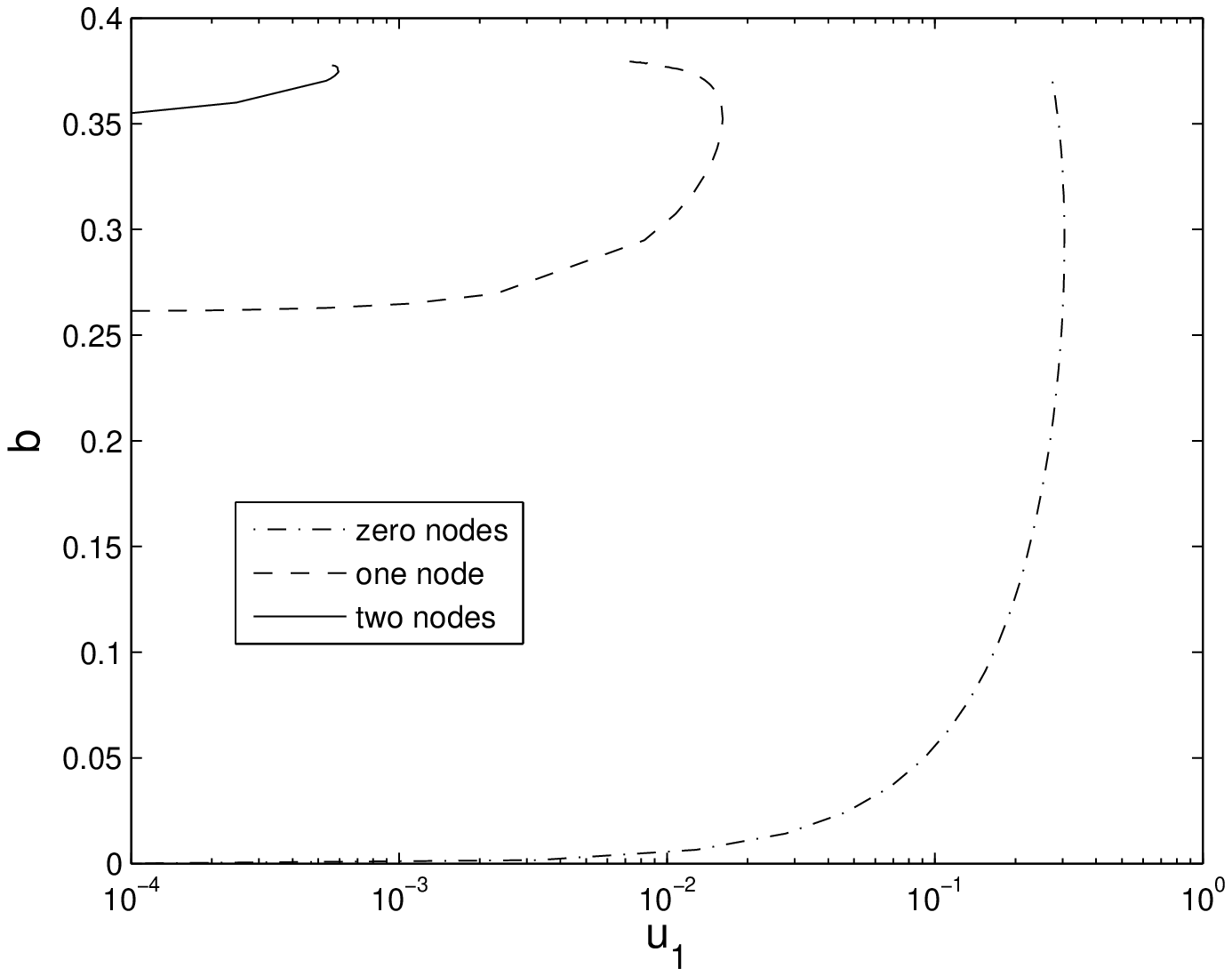}\\
\caption{\label{fig4}
The plot of initial data is given for the zero, one and two node solutions.}
\end{figure}

The relations between parameters $b$ , $u_{\infty}$ and  $u_1 > 0$
are shown in Figs. \ref{fig2}-\ref{fig4}.
Let us first discuss  Fig.\ref{fig2} where the values of $b$ are plotted as functions of the parameter
$u_{\infty}$ for the first few values of $n$. We see immediately that a branch with no node of the
function $W$ exist. In the limit $u_{\infty} \to 0$ the function $W$ converge to the constant value $W=1$
on this branch and we recover vacuum solution Eq.~(\ref{vacuum}). When $u_{\infty}$ is increasing,
$b$ clearly tends to a constant value for $u(\infty)\rightarrow u(\infty)_{cr}$.
The actual value of this constant depends on the number of nodes of the gauge
field function and increases for increasing number of nodes \cite{hartmann}. Let us point out that
when $u_{\infty}$  becomes large, the solution appears very concentrate in the region of the
origin and the numerical analysis becomes very difficult.


Complementary to Fig. \ref{fig2}, in Fig. \ref{fig3}
we show the dependence of the parameter $u_1$ on the parameter $u(\infty)$.
The parameter $u_1$ first increases with $u(\infty)$, reaches a maximal value
and then decreases and reaches a constant value for $u(\infty)\rightarrow u(\infty)_{cr}$.
Again for a given value of $u(\infty)$, the value of $u_1$ increases with
increasing number of nodes of the gauge field function.
Finally, in Fig. \ref{fig4} we summarize the pattern of the "`initial data"' of our solutions
by presenting the dependence of the parameter $b$ on $u_1$.  The three curves finish at
\bea
(b, u_1) &\approx (0.3687, 0.2763)   \ \ &{\rm for} \ \ n=0 \ \ , \ \  \nonumber \\
(b, u_1) &\approx (0.3796, 0.00707)  \ \ &{\rm for} \ \ n=1 \ \ , \ \  \nonumber \\
(b, u_1) &\approx (0.3776, 0.00056)  \ \ &{\rm for} \ \ n=2 \ \ .
\eea
On this figure we also  see in particular the following properties:
(i) For $0<b<b_1$ there exist only solutions with no nodes of $W$
having ``monopole asymptotics", $W|_{r\to\infty}=0$.
An  example of such a solution is  given by $(b,u_1)=(0.2, 0.2693)$.
(ii) For values of $b$, $b_1 \le b < b_2$ there
exist solutions with zero and one node. Fixing for example $b = 0.27$, they are characterized by the two points
$(b,u_1)=(0.27, 0.30222)$ and $(b,u_1)=
(0.27, 0.00237)$ and correspond to  solutions with monopole asymptotics having zero nodes of
$W$ and one node of $W$ respectively.
(iii) For values of $b$ such that  $b_2 \le b < b_3$ there exist
solutions with zero, one and two nodes. Taking e.g. $b=0.36$, the three points
 $(b,u_1)=(0.36, 0.28366)$, $(b,u_1)= (0.36, 0.015866)$ and $(b,u_1)=  (0.36, 0.000247)$
with the same parameter $b$ define solutions with monopole asymptotics
having respectively zero, one and two nodes of $W$.


\begin{figure}[!htb]
\centering
\leavevmode\epsfxsize=12.0cm
\epsfbox{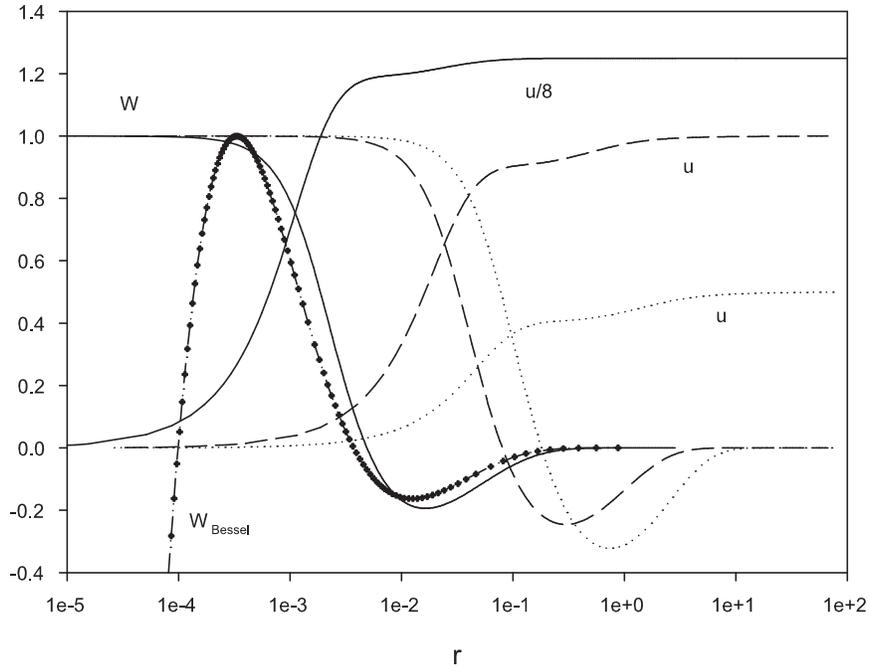}\\
\caption{
\label{fig8a}
The profiles of the gauge function $W$ and $u$
for one node solutions are presented for
$u_{\infty} = 0.5,1.0,10.0$ respectively
by the dotted, dashed and solid lines.
The Bessel function is represented by the bullets.}
\end{figure}

\begin{figure}[!htb]
\centering
\leavevmode\epsfxsize=12.0cm
\epsfbox{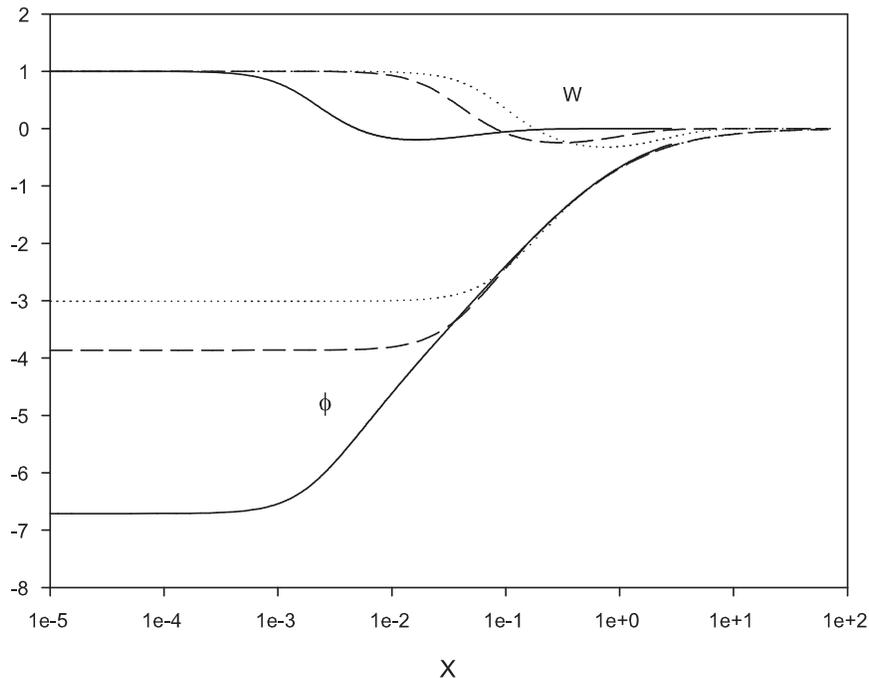}\\
\caption{
\label{fig8b}
The profiles of the gauge function $W$ and the dilaton
$\phi$ for one node solutions are presented for
$u_{\infty} = 0.5,1.0,10.0$ respectively
by the dotted, dashed and solid lines.}
\end{figure}

In order to illustrate the critical phenomenon which appears in the large $u_{\infty}$  limit
we refer to Fig.\ref{fig8a} and Fig. \ref{fig8b} where the profiles of the one-node solutions are presented
for three values of $u_{\infty}$. Several properties of the solutions can be observed on these figure.
Namely for increasing $u_{\infty}$ we observe that~: (i) the function $u(r)$ increases more and more
strongly in the region of the origin and reaches its asymptotic value when $W$ is still very close to
$W=1$; (ii) the dilaton function becomes very close to $\phi(x) = \ln(r)$ in an increasing interval of $r$
where $W(r)<<1$.
The figures further suggests that the function $W(x)$ approaches closer and closer
the profile of the Bessel function between the second extremum and infinity.

The evidence of this statement is made particularly apparent  on Fig.\ref{fig8a}
where  solutions are superposed for three different values of
$u_{\infty}$ and where the corresponding  function $W_{\rm bessel}$ is represented by the bullets line.
We see in particular that the profiles of $W$ (represented for  $u_{\infty}=0.5,1,10$ respectively by the
dotted, dashed and solid lines)  become  closer and closer to the bullets curve representing the  Bessel function
(the curve on Fig. 1 has of course to be appropriately renormalized and the argument $x$ has to be rescaled).
This rather good fit between the function $W(r)$ presenting n-nodes and the function $W_{bessel}$
breaks down in the region of the origin (the function $W_{bessel}(x)$ indeed becomes very singular for $x \to 0$).

To compute the reduced action of the solutions,
it is convenient to return to $\phi(\infty)=0$ gauge.
For small values of $u_{\infty}$ the following behaviour of the reduced action $A \equiv S_{red}$
is found
\be
A \approx 8.9 u_{\infty} \  {\rm for}  \ n=0  \ ,  \
A \approx 0.803 + 0.5 u_{\infty}  \ {\rm for}  \ n=1  \ ,  \
A \approx 0.965 + 0.1 u_{\infty}  \ {\rm for}  \ n=2  \ .
\ee
More details of the evaluation of the action for zero-node solutions are shown on Fig. \ref{fig9}.
Within our numerical accuracy, we find that $A \to 1$ in the critical limit
as its expected from the property of limiting solution Eq.~(\ref{monopole}).




\subsection{Solutions in the gauge $u(\infty)=C$}
\begin{figure}[!htb]
\centering
\leavevmode\epsfxsize=12.0cm
\epsfbox{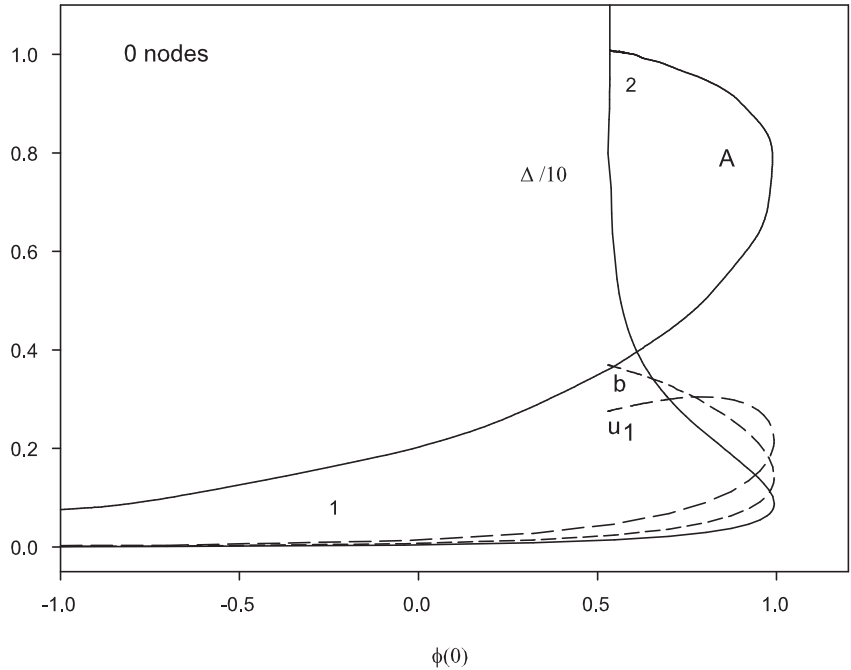}\\
\caption{
\label{fig9}
The parameters  $\Delta \equiv \phi(\infty)- \phi(0)$, $b$ and $u_1$
are plotted as functions of $\phi(0)$ for $u_{\infty}=0.2$ for zero node solutions.
Curve A represents the values of the corresponding reduced action converted to
the gauge $\phi(\infty) = 0$. }
\end{figure}

When the equations are analyzed in the `gauge' such that $\phi(\infty) = 0$ and increasing the parameter
$u_{\infty}$ by hand, the solutions become rapidly very concentrated in a region around the origin
and the numerics becomes hard, as mentioned already above.
As an alternative to understand the critical phenomenon limiting the solutions for a
maximal (but $n$-depending) value of $u(\infty)$,
we studied solution with fixed $u_{\infty}$ and increasing by hand the value $\phi(0)$
as a boundary condition. The value $\phi(\infty)$ can then be determined numerically.
Different values for $u_{\infty}$ can be connected by means of
\be
u \to  \Lambda u \ \ , \ \ r \to \frac{r}{\Lambda}  \ \ ,
\ \ \phi \to \phi - \log(\Lambda) \pkt
\ee
We will discuss in details
the results in this gauge for the solutions with zero and one nodes of $W$. We would
like to point out that no new solutions were discover in the present approach, it is rather
a complimentary way of describing the solutions obtained with the normalisation $\phi(\infty)=0$.
\\
{\bf Zero node for $W$.} \\
Let us first discuss the solution corresponding to zero node of the function $W$. If we solve the equations
with a fixed value for $u_{\infty}$ and changing the value of $\phi(0)$ by hand,
the numerical results show that when $\phi(0)$ decreases, the values $\phi(0)$ and $\phi(\infty)$
become very close to each other and that the function $\phi(x)$ becomes practically constant.
As a consequence the equations for
$W(r)$ and $u(r)$ reduce to the equations of the monopole of SU(2) gauge theory in the BPS limit.
This correspond to the first part of zero node branch labelled '1' on Fig.\ref{fig9}.
The fact that the parameter $b$ and $u_1$ tends to
zero in the limit $\phi(0) \to -\infty$ in the same limit on Figs. \ref{fig2}-\ref{fig4}
is due to the fact that the natural radial variable of the monopole solution is
rescaled by a factor $\exp(-\phi(0))$,
so that the monopole solution appears 'diluted' by the dilaton in the corresponding limit.
For small enough $\phi(0)$ this is confirmed by the numerical results which show clearly that
the profiles of $W$, $u$ hardly differ from the (suitably rescaled) BPS solution Eq.~(\ref{bps}).

\begin{figure}[!htb]
\centering
\leavevmode\epsfxsize=12.0cm
\epsfbox{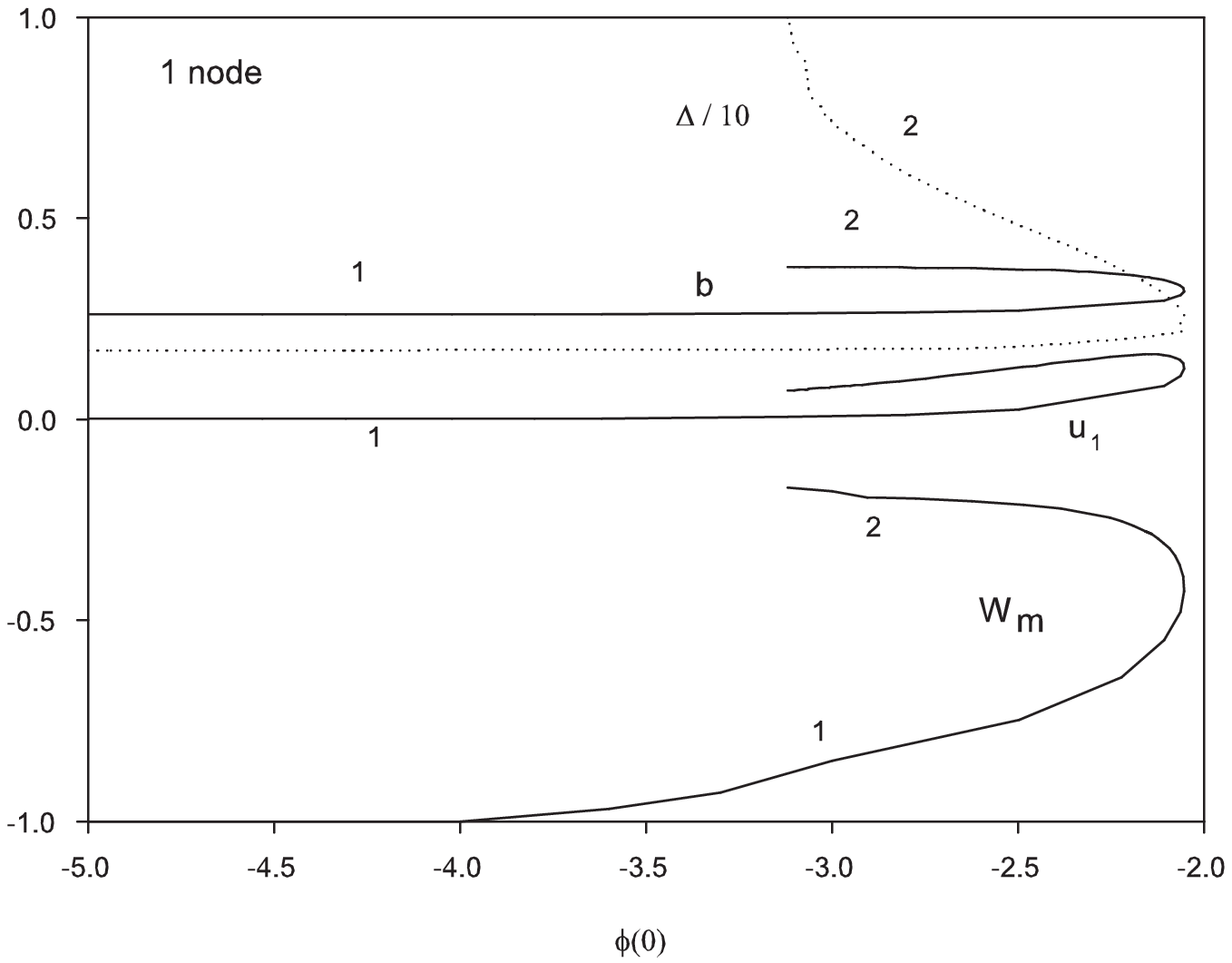}\\
\caption{
\label{fig7b}
The parameters $\Delta \equiv \phi(\infty)-\phi(0)$, $b$, $u_1$
and the minimal value $W_m$ of the gauge function
$W$ are plotted as functions of $\phi(0)$ for $u_{\infty}=0.2$
for one node solutions.}
\end{figure}

When we increase the value $\phi(0)$, it appears that solutions can be constructed
up to a maximal value of this parameter (e.g. we find $\phi(0)_m \approx 0.92$ for $u_{\infty}=0.2$)
and that a second part of this branch of solutions exists, terminating at $\phi(0) =\phi(0)_m$.
It is labelled '2' on Fig.\ref{fig9}.
When we decrease $\phi(0)$ on this part of the curve, we observe that
the parameter $b$ and the difference $\Delta \equiv \phi(\infty)-\phi(0)$ increases
while value of parameter $u_1$ decreases. However, the numerical results
strongly indicate that $\Delta$ become very large (likely infinite)
for $\phi(0) \sim 0.54$ while $b$ and $u_1$ stay finite. The numerical analysis become
very difficult in this limit.
This suggests, however the occurrence of a critical value $\phi(0)_c$ limiting the domain of existence of the
second part of this
branch in the parameter $\phi(0)$. For this choice of $\phi(0)$, the gap $\Delta$ becomes arbitrarily
large and the dilaton function is very well approximated by $\phi(r) = \ln(r)$ in the intermediate region.
\\
{\bf One node for $W$}
\\
The case of solutions with one node of the function $W$ is also instructive  and, we guess,
leads to a pattern of solutions corresponding to several nodes of $W$.
Increasing $\phi(0)$ from $\phi(0)=-\infty$, we construct the first part of this branch of
solutions corresponding to the deformation (by the  field $u(x)$) of the
first solution of the sequence of dilaton-Yang-Mills solutions constructed in \cite{lm92} .
This branch of one-node solutions can be constructed up to a maximal value of $\phi(0)$. For instance
we find $\phi(0)_{m} \approx -2.05$ where we have $u_{\infty}\approx 0.126$, $b\approx 0.318$, $\Delta \approx 2.5$,
as shown  on Fig.\ref{fig7b}. This branch is labelled  '1' on the figure.
When $\phi(0)$ decreases on this part of the curve
(which is equivalent to the limit $u_{\infty}\to 0$),
we observe that the minimum of $W$ (denoted by $W_m$ on the figure) progressively approaches $-1$
and the solution  approaches the solution
of \cite{lm92} with $W(0)=1$, $W(\infty)=-1$.

However, this is not the end of the story; indeed the second part of this branch of solutions
can be constructed for $\phi(0) < \phi(0)_{m}$. This part of the curve
is represented by the label '2' on Fig.\ref{fig7b}.
Like the case of the zero-node solution, it seems that this curve is limited by a critical value of $\phi(0)$.
Decreasing $\phi(0)$ on this part of one node branch
(which is equivalent to increase $u_{\infty}$ in reference to Figs. \ref{fig2}-\ref{fig4}),
we observe that the values of $\Delta$ and $b$ increases and value of $u_1$ decreases
and that, again, $\Delta$ seems to go to infinity
when $\phi(0)$ approaches the critical value $\phi(0)_{m}$.
As a further check of this statement that the Bessel function plays a crucial role in the critical phenomenon,
we report on Fig. \ref{fig7b} the numerical value of $W_m$ (the local minimum of $W(r)$)
 and observe that this quantity approaches the value $W_m \approx -0.16$ on the curve '2'.
 This coincides with the ratio $W(x_1)/W(x_2)$ given in  Eq.(\ref{bessel_extrema}).
\\
{\bf More nodes for $W$}
\\
For solutions with $n$ nodes of $W$, we expect that the scenario will be similar as
the two cases we have analyzed in details. In particular, we conjecture that:
(i) the second part of the curve corresponding to $n-$ node branch of solution can be constructed
up to a minimal value of the parameter $\phi(0)$;
(ii) in the large $u_{\infty}$-limit,
the gauge function $W$ of the non Abelian Euclidean solutions
will approach the profile of the Bessel function between the $n+1$-th extremum and infinity;
(iii) in this same limit $b$ will approach the absolute value of the second derivative of the function
$W_{\rm bessel}$ at the corresponding extremum.
Numerical investigation of the two-node solution supports this interpretation. A more involved  check
of this conjecture for generic $n$ would require considerable numerical efforts with our numerical technique
or a more refined analytical techniques of the type developed by D. Maison and collaborators
\cite{bfm94,bfm06}.

\section{Concluding remarks} \label{conclusions}

Motivated mainly from string theory we investigated the Euclidean version
of a model involving a Yang-Mills field coupled to a dilaton field.
We have constructed numerically different branches of finite action solutions
of the resulting equations of motion.

Euclidean theory is typically associated to tunnelling processes and to
field theory at non-zero temperature. The solutions discussed in the present paper
might be relevant to string theory at non-zero temperature. In any case
they provide new saddle points in the Euclidean path integral.

Our Euclidean model is very similar to static spherically symmetric sector of
spontaneously broken $SU(2)$ gauge theory coupled to a dilaton field,
which was investigated in \cite{fg96,fg98}
and a discrete family of regular solutions were found corresponding to
generalization of the 't Hooft-Polykov monopoles and their radial excitations.
Main difference between Euclidean theory under investigation and static sector
of spontaneously broken $SU(2)$ gauge theory is in a way how dilaton field
is coupled to electric part of the gauge field resp. Higgs field.

The Yang-Mills-dilaton theory studied in the present paper can be
viewed as a limit of vanishing gravitational constant $G\to 0$ of the
Einstein-Yang-Mills-dilaton theory investigated in \cite{br06}.
Our results might help to find new branches of solutions in
Einstein-Yang-Mills-dilaton theory.

An interesting question concerning Euclidean solutions is the number of negative modes
in the spectrum of small perturbations about these solutions.
It is known that the instanton solutions,
which describe mixing between equal energy states
have no negative modes (at most have zero modes). Bounce solutions,
which describe decay of a metastable vacuum have exactly one negative
mode \cite{col88,klt00,lav00,grtu01}.
Whereas presence of more then one negative mode typically shows that
there are other solutions with lower action \cite{lav06}.
For better understanding of the role of solutions which we discussed here and
for their interpretation it is important to study question of negative modes about them.
We plan to come to this topic in further investigation.

\section*{Acknowledgements}
We  are grateful to Dr. Betti Hartmann for numerous
discussions at the initial stages of this work and in particular for communicating
us some unpublished results on the topic.
The main part of G.L.'s work has been done during his visit to Switzerland.
He would like to thank the theory group of Geneva University and especially Prof. Ruth Durrer
for kind hospitality and wishes to
acknowledge financial support of the Tomalla foundation and SNF.


\end{document}